# User-Aware Power Management for Mobile Devices

Geunsik Lim[1], Changwoo Min[2], Dong Hyun Kang[3], and Young Ik Eom[4]
Sungkyunkwan University, Republic of Korea[1234]   Samsung Electronics, Republic of Korea[12]
{leemgs[1], multics69[2], kkangsu[3], yieom[4]}@skku.edu, {geunsik.lim[1], changwoo.min[2]}@samsung.com

*Abstract*—The power management techniques to extend battery lifespan is becoming increasingly important due to longer user applications' running time in mobile devices. Even when users do not use any applications, battery lifespan decreases continually. It occurs because of service daemons of mobile platform and network-based data synchronization operations. In this paper, we propose a new power management system that recognizes the idle time of the device to reduce the battery consumption of mobile devices.

## I. INTRODUCTION

The physical extension of battery size impairs the mobility of mobile devices due to the increased battery weight and volume. In addition, the increase in CPU utilization accelerates the battery consumption. Therefore, the battery management [1], [2] becomes increasingly important in the mobile environment. Especially, the power usage of peripheral devices during users' idle time increases the battery consumption drastically. The existing power management techniques made an effort to settle this issue with hardware development. [3] However, this approach has its limitation that damages microminiaturization of mobile devices. Therefore, the system-wide software technique for suppressing the unnecessary battery consumption is always required. In this paper, we assert that one of the major power consumptions of mobile devices happen because of the operations to keep the mobile system running all the time regardless of the unused time of user device. In the next section, we depict a timer-based user-aware power management scheme that suppresses the inevitable power consumption of mobile platforms while the users do not use the mobile device.

## II. PROPOSAL FOR EXTENDING BATTERY LIFESPAN

A motivating example that quantifies the impact of mobile platform pollution on battery lifespan can be seen in Figure 1. It depicts the typical pattern of power consumption of a mobile device without running any user applications (*factory reset*). Figure 1 shows breakdown of the power consumption that consumes 19% (315 *mAh*) of the battery while user does not use the mobile device for eight hours on *dual-core smart phone* (1650 *mAh*). Using a battery monitoring unit (BMU), we classify the sources of power consumption in a mobile device. The y-axis in Figure 1 represents the power usage. The maximum power of the y-axis is 315 *mAh*. The most power is consumed in (1) device operations & service daemons (e.g. cell standby, device idle, and Android OS) and (2) application daemons (e.g. Wi-Fi, Screen, and Gmail). From our experiment, we verified that device operations & service daemons deplete 80% (252 *mAh*) of the battery power and application daemons deplete 20% (63 *mAh*) of the battery power.

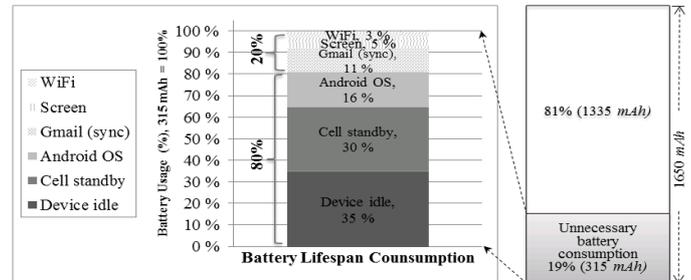

Fig. 1. The breakdown of power consumption in a mobile device under no user applications running. The y-axis 100% means that 315 *mAh* are consumed during eight hours (idle time of user device). The *mAh* refers to "milliampere hour".

However, the previous studies on power management cover the power monitoring methods on applications or devices. Most of the software approaches terminate unimportant service applications to reduce power consumption [1], [2] whenever the battery lifespan reaches at the point less than or equal to the specified threshold. That is, the existing research does not handle power consumption of the mobile platform. Therefore, we propose a new battery management system to settle the problem of power consumption shown in Figure 1. We introduce an automatic battery manager that completely controls the power supply of the mobile device based on hardware timer to suppress the battery consumption of mobile platform while user does not use the mobile device. The proposed battery manager consists of four major components as shown in Figure 2: *(1) user-space client, (2) sleep time manager, (3) sleep level controller,* and *(4) battery timer*. The *user-space client* sends user-aware time information to the *sleep time manager* in kernel-space.

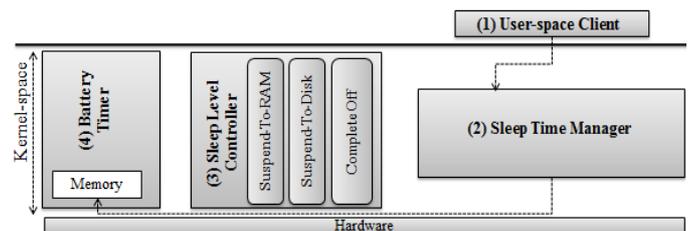

Fig. 2. The overall diagram of the proposed system

### A. User-space client

Figure 2 shows the sequential diagram when the user gives input with the power consumption policy using the *user-space client*. When the user cannot charge the battery, the user can save sleep time and wake-up time easily with the *user-space client* as a user input interface. The sleep time and the wake-up time can be decided by user regardless of the available battery capacity.

## B. Sleep time manager

The *sleep time manager* in Figure 2 manages the execution time such as sleep time and wake-up. It restarts all service daemons of a mobile device automatically when the current time of mobile device reaches a specified wake-up time. The *sleep time manager* requests actual works to battery timer. If users set sleep time of the mobile platform and wake-up time of all services with user-space client, the *sleep time manager* schedules with the time data given by user.

## C. Sleep level controller

The *sleep level controller* in Figure 2 consists of three controllers: *suspend-to-ram*, *suspend-to-disk*, and *complete off*. First, *suspend-to-ram* supplies the battery power to memory only to retain the content in memory after saving the all contents into volatile memory. This controller is effective when a user wants to run the minimal functions such as phone application and short message service (SMS) application. The minimal functions can be chosen by the user. Second, the *suspend-to-disk* preserves all the contents into storage using swap mechanism to avoid the power supply of memory additionally against the *suspend-to-ram*. The *suspend-to-disk* is more energy-efficient than the *suspend-to-ram*. However, the engineering cost of the *suspend-to-ram* is less than that of the *suspend-to-disk* due to the device snapshot and the device restoration [4] in real environment. Finally, the *complete-off* suppresses the battery supply completely until all services of the mobile device restart to execute user manipulation. The *complete-off* and the *suspend-to-disk* only need battery timer in implementing the sleep procedure of the schemes.

## D. Battery timer

In a state where the supply of the battery is cut off, a hardware level timer restores all services to the runnable status. When the proposed system suppresses the power supply of mobile devices and all services at a specified time instantly, the *battery timer* in Figure 2 executes automatically and re-launches all services after a specified time. At this time, the battery timer supplies the power to the mobile device using timer-based *clocksource* and *clockevent*. The *battery timer* executes sleep procedure and wake-up procedure by hardware timer, periodically. The *battery timer* saves time information into memory of hardware timer. The proposed approach depends on the hardware design because the *battery timer* operates based on hardware timer for implementation in real device. The implementation of the *battery timer* is easy because mobile devices operate alarm software based on hardware timer called *real time clock*. In the state where the power supply of the mobile device is interrupted, the *battery timer* preserves the time data transmitted from the *sleep time manager* for the wake-up procedure.

## III. EXPERIMENTAL RESULTS

To experiment our approach in various environments, we experimented the proposed system in four mobile devices: *dual-core smart phone (1650 mAh)*, *quad-core smart phone (2100 mAh)*, *quad-core tablet (4325 mAh)*, and *laptop (4400 mAh)*. We compared the power consumption of battery results in the mobile platform without running any user applications (*factory reset*) after testing 8 hours as a long idle time of the user.

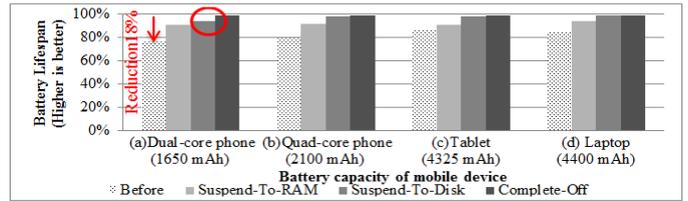

**Fig. 3. Battery consumption of mobile platform**

Figure 3 shows the comparison result of the power consumption between the existing system (before) and the proposed systems in three configurations (e.g. *suspend-to-ram*, *suspend-to-disk*, and *complete-off*) in our mobile device without running any user applications. The value of the x-axis means the different battery capacity of devices. The value of the y-axis means the available battery capacity. From our experimental result, our system drastically extended the battery lifespan up to 18% against the existing system without any user applications. We got the reasonable results at the four mobile devices as shown in Figure 3. Rate of power consumption of *suspend-to-ram* is faster than that of *suspend-to-disk*. Rate of power consumption of *suspend-to-disk* is similar to that of *complete-off* due to only the power supply of the *battery timer*. However, in case (a), the small circle in Figure 3 shows that power consumption of *suspend-to-disk* is exhausted more than that of *complete-off*. From our analysis, we found out that these results were happened due to the power supply of peripheral devices that depends on system-on-chip design.

## IV. CONCLUSION

In this paper, we propose a new power management system that controls the power supply completely based on hardware timer. Our system manages the battery power consumption via four major components such as *battery timer*, *sleep level controller*, *sleep time manager*, and *user-space client*. We showed our system extends the battery lifespan up to maximum 18% compared to the existing system. In addition, the proposed system reschedules automatically all services in advance before the user tries to reuse the mobile device.

ACKNOWLEDGMENT

This research was supported by Basic Science Research Program through the National Research Foundation of Korea (NRF) funded by the Ministry of Education, Science and Technology (2010-0022570). We thank Jeehong Kim and Sangyong Yang for their feedback and comments.